\documentstyle[aps,prb,epsf]{revtex}
\begin{document}
\title{On Local Symmetric Order Parameters of Vortex Lattice States}
\author{Masa-aki~ {\sc Ozaki}\cite{ozaki}, Yoshiki~ {\sc Hori}\cite{hori} 
and Akira~ {\sc Goto}\cite{goto}\\ \bigskip}
\address{$^*$Department of Physics, Kochi University, Kochi 780-8072, Japan}
\address{$^{\dagger , \ddagger}$Kochi National College of Technology, 
Nankoku 783-8505, Japan\\ \bigskip}
\newcommand{\Db}{\bbox{D}}
\newcommand{\Bb}{\bbox{B}}
\newcommand{\Lb}{\bbox{L}}
\newcommand{\Lbb}{\bbox{L}}
\newcommand{\Dbb}{\bbox{D}}
\newcommand{\Cbb}{\bbox{C}}
\newcommand{\Sbb}{\bbox{S}}
\date{\today}
\maketitle
\begin{abstract}
This paper gives a  new refined definition of local symmetric order parameters 
(OPs)($s$-wave, $d$-wave and $p$-wave order parameters) of vortex lattice 
states for  singlet superconductivity.$s$-wave, $d$-wave and $p$-wave OPs 
at a site $(m,n)$ are defined as $A$, $B$ and $E$ representations 
of the four fold rotation ${\Cbb }_4$ at the site $(m,n)$
of nearest neighbor OPs 
$\langle a_{(m,n)\downarrow}a_{(m+1,n)\uparrow}\rangle$ etc.
The new OPs have  a  well defined nature such that an OP(e.g. $d$-wave) 
at the site obtained under translation by a lattice vector 
(of the vortex lattice) from a site $(m,n)$ is expressed by the corresponding 
OP (e.g. $d$-wave)  at the site $(m,n)$ times a phase factor.
The  winding numbers of  $s$-wave and $d$-wave OPs are given.
\end{abstract}

\bigskip\bigskip
\newcommand{\kb}{\bbox{k}}
\newcommand{\rb}{\bbox{r}}
\newcommand{\qb}{\bf q}
\newcommand{\eb}{\bbox{e}}
\newcommand{\Sb }{\bbox{S}}
\newcommand{\kr}{\rm k}
\newcommand{\lb}{\bbox{l}}
\newcommand{\qr}{\rm q}
\newcommand{\Cb}{\bbox{C}}
\newcommand{\Ab}{\bbox{A}}
\newcommand{\Gr}{\rm G}
\newcommand{\Xb}{\bf X}
\newcommand{\Tb}{\bbox{T}}
\newcommand{\Ar}{\rm A}
\newcommand{\Br}{\rm B}
\newcommand{\Pb }{\bbox{P}}
\newcommand{\Er}{\rm E}
\newcommand{\sigmab}{\bbox{\sigma}}
\newcommand{\Phib}{\bbox{\Phi}}
\def\@cite#1#2{$^{\hbox{\scriptsize{#1\if@tempswa , #2\fi}}}$}

Growing interest has been focused on microscopic works on the basis of the 
Bogoliubov de Gennes (BG) equation  of  an isolated vortex\cite{rf:Hayashi1} 
or a vortex lattice\cite{rf:Kita1} in both conventional and unconventional 
superconductors. In a previous paper\cite{rf:Ozaki1}(referred to as I)
we gave a  group theoretical classification of vortex lattice solutions 
of the BG equation in a uniform magnetic field of the extended Hubbard model 
on a two-dimensional square lattice. In that paper, we considerd conventional 
definitions\cite{rf:Soininen,rf:MacDonald,rf:Tanaka}
of $s$-wave, $d$-wave and $p$-wave local symmetric order parameters (OPs),
which are valid  definitions of OPs in a spatially uniform superconducting 
state in the absence of a magnetic field.
We  showed  that, e.g., $s$-wave OP at  the site obtained under translation 
by a lattice vector (of the vortex lattice) from a site $(m,n)$
is always a linear combination of   $s$-wave, $d$-wave and $p$-wave 
OPs at the site  $(m,n)$.
 We then pointed out that  it is not appropriate to assign conventional 
 order parameters at each site.

In this paper  we give more  appropriate definitions of OPs such that an  OP 
at the translated site is expressed by the corresponding OP at the original 
site times a phase factor. The model and notation used here are the same as
those in I.

We consider singlet superconductivity on a square lattice with inter-atomic 
spacing $a$. The Hamiltonian for the system is given by
\begin{eqnarray}
  H&=&-t\sum_{(m,n)s}\{ (e^{iKn}a^{\dag }_{(m,n)s}a_{(m+1,n)s}
+e^{-iKm}a^{\dag }_{(m,n)s}a_{(m,n+1)s})+{\rm H.c}\} \nonumber \\
& &-\mu \sum_{(m,n)s}a^{\dag }_{(m,n)s}a_{(m,n)s}\nonumber \\
& &+U\sum_{(m,n)}a^{\dag }_{(m,n)\uparrow }a_{(m,n)\uparrow }
a^{\dag }_{(m,n)\downarrow }a_{(m,n)\downarrow }\nonumber \\
& &+V\sum_{(m,n)ss^{\prime }}\{ a^{\dag }_{(m,n)s}a_{(m,n)s}
a^{\dag }_{(m+1,n)s^{\prime}}a_{(m+1,n)s^{\prime }}\nonumber \\
& &
\label{t Hamiltonian}
+a^{\dag }_{(m,n)s}a_{(m,n)s}
a^{\dag }_{(m,n+1)s^{\prime}}a_{(m,n+1)s^{\prime }}\},
\end{eqnarray} 
where $(m,n)$ denotes a site in a square crystal lattice, 
${\Bb }=(0,0,B)$ is a magnetic field, $K=(\pi \phi )/(2\phi _0)$,
$\phi =Ba^2$ is the  magnetic flux through a unit cell of the crystal lattice, 
and  $\phi _0=ch/2e$ is the flux quantum. 
The symmetry group of the system is given by
\begin{equation}
G_0=(e+tC_{2x}){\Cb }_4{\Tb }{\Sb }{\Phib },
\end{equation}
where ${\Tb }$ is the group of the magnetic translation,\cite{rf:Brown}
consisting of the elements $T(Ma{\eb }_x+Na{\eb }_y)$ ($M,N=$integer ),  
such that
\newpage
\begin{eqnarray}
\label{eq:magtra}
T(Ma{\eb }_x+Na{\eb }_y)\cdot a^{\dag}_{(m,n)s}
&=&e^{iK(Mn-Nm)}a^{\dag }_{(m+M,n+N)s}
\end{eqnarray}
and ${\Cb }_4=(e,C^+_{4z},C_{2z},C^{-}_{4z})$ is 
the four-fold rotation group around the origin $(0,0)$. 
Its generator's action on the Fermion operator is given by
\begin{eqnarray}
\label{C4z}
C^+_{4z}\cdot a^{\dag }_{(m,n)s}&=&a^{\dag }_{(-n,m)s}.
\end{eqnarray}
The action of $tC_{2x}$ ($t$ is  time reversal)\cite{rf:Ozaki1} on 
the Fermion operator  is 
\begin{eqnarray}
tC_{2x}\cdot (fa^{\dag }_{(m,n)\uparrow})&=&
-f^*a^{\dag }_{(m,-n)\downarrow },\nonumber \\ 
\label{C2x}
tC_{2x}\cdot (fa^{\dag }_{(m,n)\downarrow })&=&
f^*a^{\dag }_{(m,-n)\uparrow },
\end{eqnarray}
where $f$ is a complex number. Also in Eq. (2),
${\Sb }$ is the group of the spin rotation ($SU(2)$), and ${\Phib }$ is the 
group of the global gauge transformation.  

Hereafter, for illustrative purposes, we restrict our consideration to 
the symmorphic tetragonal vortex lattice states for the case 
$\phi =\phi _0/p^2(p:{\rm integer })$, in which the center point of each 
rotation is at a site of the crystal lattice.
In this case there are four types of tetragonal vortex lattice solutions 
of the BG equation characterized by invariance groups\cite{rf:Ozaki1}: 
\begin{eqnarray}
\label{Gl}
G_{(l)}&=&(e+tC_{2x})\widetilde{\Cb }^l{\Lb }{\Sb}\ \ (l=0,2,\pm 1).
\end{eqnarray}
Here 
${\Lb }$ is a subgroup of ${\Tb }{\Phib }$ consisting elements 
$L(Mp{\eb }_x+Np{\eb }_y)$ such that
\begin{eqnarray}
L(Mpa{\eb }_x+Npa{\eb }_y)a^{\dag }_{(m,n)s}&\equiv &e^{i\frac{\pi }{2}(MN+M+N)}
T( Mpa{\eb }_x+Npa{\eb }_y)a^{\dag }_{(m,n)s},
\end{eqnarray}
and 
\begin{eqnarray}
\widetilde{\Cb }^l&=&\sum_j e^{i\frac{2\pi}{4}lj}C^j_{4z},
\end{eqnarray}
where the  phase factor $e^{i\frac{2\pi}{4}lj}$ should be understood to appear 
when a group element acts on the space $W(a^{\dag }a^{\dag })$.
(The case $l=-1$ is overlooked in I.)
From ${\Lb }$ and $\widetilde{\Cb }^l$ invariance of the generalized
density matrix we have 
\begin{eqnarray}
\langle (L(Mpa{\eb }_x+Npa{\eb }_y)a_{(m,n)s})
(L(Mpa{\eb }_x&+&Npa{\eb }_y)a_{(m^{\prime }n^{\prime })s^{\prime }})
\rangle \nonumber \\
&=&\langle a_{(m,n)s}a_{(m^{\prime },n^{\prime })s^{\prime }}\rangle
\nonumber \\
\langle e^{-i\frac{2\pi}{4}l}(C_{4z}^+a_{(m,n)s})
(C_{4z}^+a_{(m^{\prime }n^{\prime })s^{\prime }})\rangle
&=&\langle a_{(m,n)s}a_{(m^{\prime },n^{\prime })s^{\prime }}\rangle
\end{eqnarray}
Then we have 
\begin{eqnarray}
\langle a_{(m+Mp,n+Np)s}&a&_{(m^{\prime }+Mp,n^{\prime }+Np)s^{\prime }}
\rangle \nonumber \\
&=&e^{i\pi (MN+M+N)+i\frac{\pi }{2p}(M(n+n^{\prime })-N(m+m^{\prime }))}
\langle a_{(m,n)s}a_{(m^{\prime },n^{\prime })s^{\prime }}\rangle,\nonumber \\
\label{OPMN}
\langle a_{(-n,m)s}a_{(-n^{\prime },m^{\prime })s^{\prime }}\rangle 
&=&e^{i\frac{\pi }{2}l}
\langle a_{(m,n)s}a_{(m^{\prime },n^{\prime })s^{\prime }}\rangle.
\end{eqnarray}
For the case $l=-1,p=3$, using (\ref{OPMN}) we obtain a schematic pattern of 
bond OPs such as $\langle{a_{(m,n)\downarrow}a_{(m+1,n)\uparrow}}\rangle$, 
which is displayed in Fig. 1.
\newpage
\begin{figure}
\epsfysize= 6cm
\centerline{\epsfbox{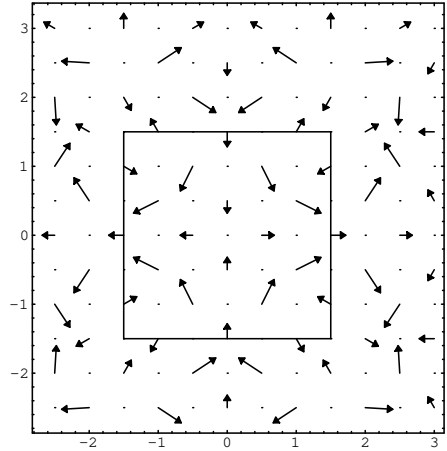}}
\caption{Schematic pattern of bond OP $\langle a_{(m,n)\downarrow}
a_{(m+1,n)\uparrow}\rangle $ and 
$\langle a_{(m,n)\downarrow}a_{(m,n+1)\uparrow}\rangle$ for $p=3,\ l=-1$. 
The angle of an arrow with respect to the horizontal $x$ axis  represents
the phase of the OP. The inner square represents a magnetic unit cell.}
\label{fig:1}
\end{figure}
\noindent
It is known\cite{rf:Soininen,rf:Vol} that even in the case of 
an isolated  vortex line, a $d$-wave OP induces an $s$-wave OP.
In I, we use conventional $s$-wave,$d$-wave and $p$-wave 
OPs\cite{rf:Soininen,rf:MacDonald,rf:Tanaka} on a site $(m,n)$ as follows:
\begin{eqnarray}
S(m,n)&=&\frac{1}{4}\{ {\langle }a_{(m,n)\downarrow }a_{(m+1,n)\uparrow
}{\rangle }
+{\langle }a_{(m,n)\downarrow }a_{(m-1,n)\uparrow }{\rangle }
\nonumber \\
& & \ \ +{\langle }a_{(m,n)\downarrow }a_{(m,n+1)\uparrow }{\rangle }
+{\langle }a_{(m,n)\downarrow }a_{(m,n-1)\uparrow }{\rangle }\},
\nonumber \\
D(m,n)&=&\frac{1}{4}\{ {\langle }a_{(m,n)\downarrow }a_{(m+1,n)\uparrow
}{\rangle }
+{\langle }a_{(m,n)\downarrow }a_{(m-1,n)\uparrow }{\rangle }
\nonumber \\
& & \ \ -{\langle }a_{(m,n)\downarrow }a_{(m,n+1)\uparrow }{\rangle }
-{\langle }a_{(m,n)\downarrow }a_{(m,n-1)\uparrow }{\rangle }\},
\nonumber \\
P_x(m,n)&=&\frac{1}{2}\{ {\langle }a_{(m,n)\downarrow }a_{(m+1,n)\uparrow
}{\rangle }
-{\langle }a_{(m,n)\downarrow }a_{(m-1,n)\uparrow }{\rangle }\},
\nonumber \\
\label{SDP OP}
P_y(m,n)&=&\frac{1}{2}\{ {\langle }a_{(m,n)\downarrow }a_{(m,n+1)\uparrow
}{\rangle }
-{\langle }a_{(m,n)\downarrow }a_{(m,n-1)\uparrow }{\rangle }\}.
\end{eqnarray}    
Then we have shown that these  OPs  at the site $(m+Mp,n+Np)$ are expressed 
by a linear combination of $s$-wave, $d$-wave and $p$-wave OPs at the site 
$(m,n)$: 
\begin{eqnarray}
D(m+Mp,n+Np)
&=&\frac{1}{2}e^{i\pi(MN+M+N)}e^{i\frac{\pi}{p}(Mn-Nm)}\nonumber \\
& &\times \{ (\cos (\frac{\pi N}{2p})-\cos (\frac{\pi M}{2p}))S(m,n)\nonumber \\
& &+(\cos (\frac{\pi N}{2p})+\cos (\frac{\pi M}{2p}))D(m,n)\nonumber \\
& &+i\sin (\frac{\pi N}{2p})P_x(m,n)
-i\sin (\frac{\pi M}{2p})P_y(m,n)\},
\end{eqnarray}
and so on.
Then it is very difficult to recognize  the symmetry of  magnetic translation 
of the vortex lattice by inspecting  $d$ wave  OPs $D(m,n)$ and so on.
Thus  it is not appropriate to describe a vortex lattice state
by assigning  the conventional $d$-wave OP at each site. 

Here we give more refined definitions of $s$-wave, $d$-wave and $p-$wave 
OPs on a site $(m,n)$.
As shown in I, $2\pi /4$, $\pi $ and $(-2\pi) /4$ rotation around $(m,n)$
(we denote them by
$C^+_{4z}(m,n)$, $C_{2z}(m,n)$ and  $C^-_{4z}(m,n)$)
are given by 
\begin{eqnarray}
C^+_{4z}(m,n)&=&T(ma{\eb }_x+na{\eb }_y)C^+_{4z}
T(-ma{\eb }_x-na{\eb }_y),\nonumber \\
C_{2z}(m,n)&=&T(ma{\eb }_x+na{\eb }_y)C_{2z}
T(-ma{\eb }_x-na{\eb }_y),\nonumber \\
C^-_{4z}(m,n)&=&T(ma{\eb }_x+na{\eb }_y)C^-_{4z}
T(-ma{\eb }_x-na{\eb }_y)
\end{eqnarray}
\newpage
Thus we have 
\begin{eqnarray}
C^+_{4z}(m,n)\cdot a_{(m+1,n)s}&=&e^{-iK(m+n)}a_{(m,n+1)s}, \nonumber \\
C_{2z}(m,n)\cdot a_{(m+1,n)s}&=&e^{-2iKn}a_{(m-1,n)s},\nonumber \\
C^-_{4z}(m,n)\cdot a_{(m+1,n)s}&=&e^{-iK(n-m)}a_{(m,n-1)s},\nonumber \\
C^+_{4z}(m,n)\cdot a_{(m,n)s}&=&C_{2z}(m,n)\cdot a_{(m,n)s}
=C^-_{4z}(m,n)\cdot a_{(m,n)s}
\nonumber \\
\label{mnrot}
&=&a_{(m,n)s}.
\end{eqnarray}

Here we define $s$-wave, $d$-wave and $p$-wave OPs on site$(m,n)$ 
as $A, B$ and $E$ representation of the four fold rotation group 
${\Cb }_4(m,n)$ around the point $(m,n)$ consisting of \{$e$, 
$C^+_{4z}(m,n)$, $C_{2z}(m,n)$ and  $C^-_{4z}(m,n)$\} as follows :
\begin{eqnarray}
\widetilde{S}(m,n)&=&\frac{1}{4}e^{iKn}\{
\langle a_{(m,n)\downarrow}a_{(m+1,n)\uparrow}\rangle
+\langle (C^+_{4z}(m,n)a_{(m,n)\downarrow})
(C^+_{4z}(m,n)a_{(m+1,n)\uparrow})\rangle \nonumber \\
& & \ +\langle (C_{2z}(m,n)a_{(m,n)\downarrow})
(C_{2z}(m,n)a_{(m+1,n)\uparrow})\rangle \nonumber \\
& & \ +\langle (C^-_{4z}(m,n)a_{(m,n)\downarrow})
(C^-_{4z}(m,n)a_{(m+1,n)\uparrow})\rangle \}, \nonumber \\
\widetilde{D}(m,n)&=&\frac{1}{4}e^{iKn}\{
\langle a_{(m,n)\downarrow}a_{(m+1,n)\uparrow}\rangle
-\langle (C^+_{4z}(m,n)a_{(m,n)\downarrow})
(C^+_{4z}(m,n)a_{(m+1,n)\uparrow})\rangle \nonumber \\
& & \ +\langle (C_{2z}(m,n)a_{(m,n)\downarrow})
(C_{2z}(m,n)a_{(m+1,n)\uparrow})\rangle \nonumber \\
& & \ -\langle (C^-_{4z}(m,n)a_{(m,n)\downarrow})
(C^-_{4z}(m,n)a_{(m+1,n)\uparrow})\rangle \}, \nonumber \\
\widetilde{P_x}(m,n)&=&\frac{1}{2}e^{iKn}\{
\langle a_{(m,n)\downarrow}a_{(m+1,n)\uparrow}\rangle
-\langle (C_{2z}(m,n)a_{(m,n)\downarrow})
(C_{2z}(m,n)a_{(m+1,n)\uparrow})\rangle \},\nonumber \\
\widetilde{P_y}(m,n)&=&\frac{1}{2} e^{iKn}\{
\langle ( C^+_{4z}(m,n)a_{(m,n)\downarrow})
(C^+_{4z}(m,n)a_{(m+1,n)\uparrow})\rangle \nonumber \\
& & -\langle (C^-_{4z}(m,n)a_{(m,n)\downarrow})
(C^-_{4z}(m,n)a_{(m+1,n)\uparrow})\rangle \}.
\end{eqnarray}
Here the factor $e^{iKn}$ is used to obtain  a symmetrical expression. 
Using (\ref{mnrot}) we have
\begin{eqnarray}
\widetilde{S}(m,n)&=&\frac{1}{4}\{
e^{iKn}\langle a_{(m,n)\downarrow}a_{(m+1,n)\uparrow}\rangle
+e^{-iKm}\langle a_{(m,n)\downarrow }a_{(m,n+1)\uparrow }\rangle \nonumber \\
& & +e^{-iKn}\langle a_{(m,n)\downarrow }a_{(m-1,n)\uparrow }\rangle
+e^{iKm}\langle a_{(m,n)\downarrow }a_{(m,n-1)\uparrow }\rangle \}, \nonumber \\
\widetilde{D}(m,n)&=&\frac{1}{4}\{
e^{iKn}\langle a_{(m,n)\downarrow}a_{(m+1,n)\uparrow}\rangle
-e^{-iKm}\langle a_{(m,n)\downarrow }a_{(m,n+1)\uparrow }\rangle \nonumber \\
& & +e^{-iKn}\langle a_{(m,n)\downarrow }a_{(m-1,n)\uparrow }\rangle
-e^{iKm}\langle a_{(m,n)\downarrow }a_{(m,n-1)\uparrow }\rangle \}, \nonumber \\
\widetilde{P_x}(m,n)&=&\frac{1}{2}\{
e^{iKn}\langle a_{(m,n)\downarrow}a_{(m+1,n)\uparrow}\rangle
-e^{-iKn}\langle a_{(m,n)\downarrow }a_{(m-1,n)\uparrow }\rangle, \nonumber \\
\widetilde{P_y}(m,n)&=&\frac{1}{2} \{
e^{-iKm}\langle a_{(m,n)\downarrow }a_{(m,n+1)\uparrow }\rangle \nonumber \\
& & 
-e^{iKm}\langle a_{(m,n)\downarrow }a_{(m,n-1)\uparrow }\rangle \}
\end{eqnarray}
From (\ref{OPMN}) we can see  that
\begin{eqnarray}
\widetilde{S}(m+Mp,n+Np)&=&e^{i\pi (MN+M+N)}e^{i\frac{\pi }{p}(Mn-Nm)}
\widetilde{S}(m,n),\nonumber \\
\widetilde{D}(m+Mp,n+Np)&=&e^{i\pi (MN+M+N)}e^{i\frac{\pi }{p}(Mn-Nm)}
\widetilde{D}(m,n),\nonumber \\
\widetilde{P_x}(m+Mp,n+Np)&=&e^{i\pi (MN+M+N)}e^{i\frac{\pi }{p}(Mn-Nm)}
\widetilde{P_x}(m,n),\nonumber \\
\widetilde{P_y}(m+Mp,n+Np)&=&e^{i\pi (MN+M+N)}e^{i\frac{\pi }{p}(Mn-Nm)}
\widetilde{P_y}(m,n) \nonumber \\
\widetilde{S}(-n,m)&=&
e^{i\frac{\pi }{2}l}\widetilde{S}(m,n),\nonumber \\
\widetilde{D}(-n,m)&=&-
e^{i\frac{\pi }{2}l}\widetilde{D}(m,n),\nonumber \\
\widetilde{P_x}(-n,m)&=&-
e^{i\frac{\pi }{2}l}\widetilde{P_y}(m,n),\nonumber \\
\label{OPROT2}
\widetilde{P_y}(-n,m)&=&
e^{i\frac{\pi }{2}l}\widetilde{P_x}(m,n)
\end{eqnarray}
\newpage
Thus these new OPs do not mix due to the invariance magnetic 
translation ${\Lb }$.
In Figs. 2 and 3 we show schematic patterns of $\widetilde{S}(m,n)$ and 
$\widetilde{D}(m,n)$ corresponding  to Fig. 1 for  the case $l=-1,p=3$.\\
\begin{minipage}[t]{0.47\textwidth}
\begin{figure}
\epsfysize=5cm
\centerline{\epsfbox{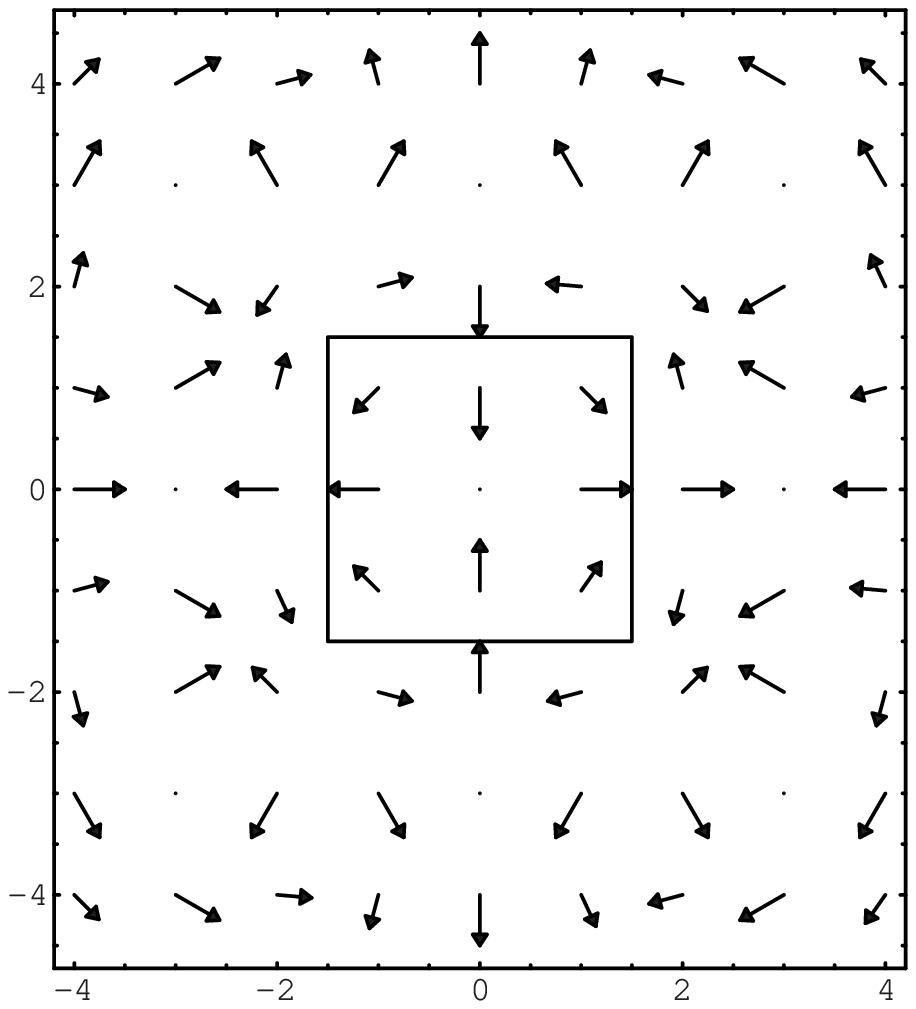}}
\caption{A schematic pattern of $\widetilde{S}(m,n)$ for the case $p=3,\ 
l=-1$.}
\end{figure}
\end{minipage}
\hfill
\begin{minipage}[t]{0.47\textwidth}
\begin{figure}
\epsfysize=5cm
\centerline{\epsfbox{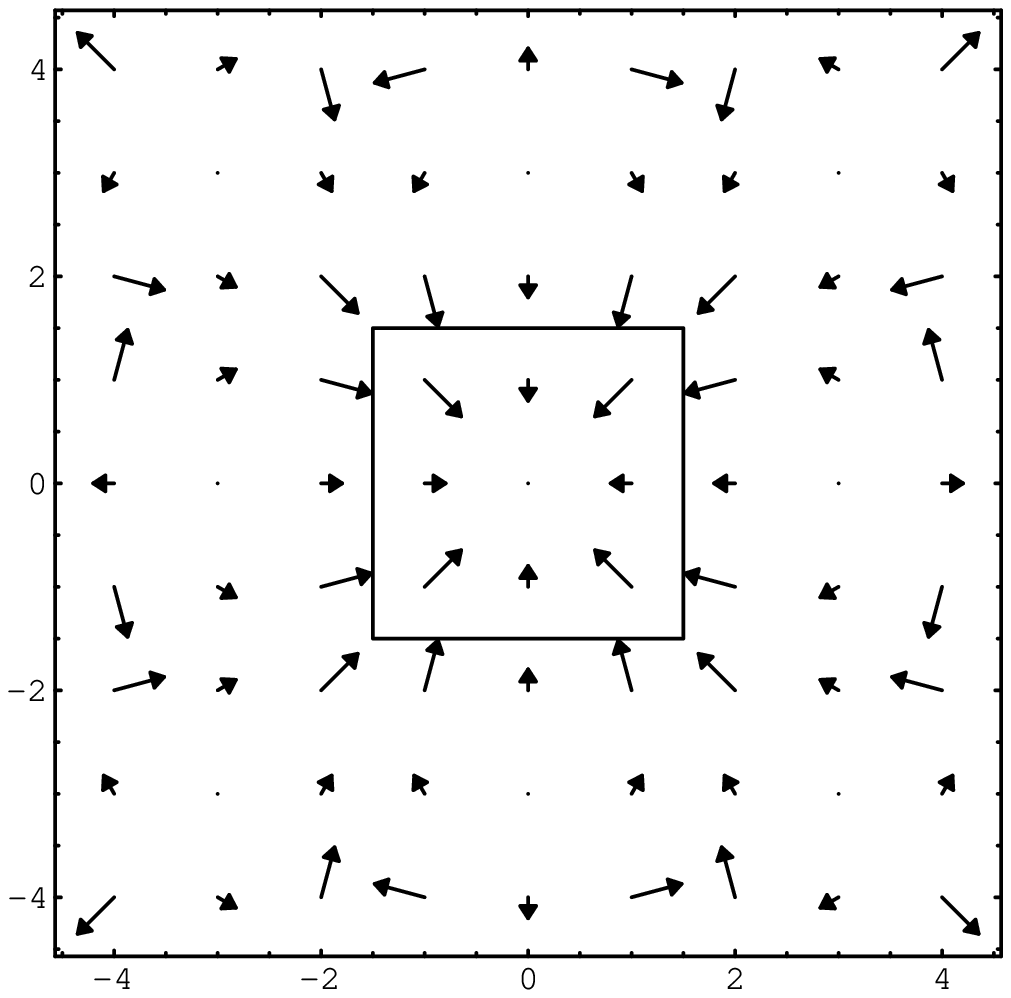}}
\caption{A schematic pattern of $\widetilde{D}(m,n)$ for the case $p=3,\ 
l=-1$.}
\end{figure}
\end{minipage}
\bigskip

In the process of making one rotation around the origin $(0,0)$:
$(m,n)\Rightarrow (-n,m) \Rightarrow (-m,-n) \Rightarrow (n,-m) 
\Rightarrow (n,m)$,
total change of the phase of the $s$-wave OP is $2\pi l$. 
The corresponding total change \\ 
\begin{minipage}[t]{6.5cm}
\begin{table}
\caption{Winding number $N_W$ aroud $(0,0)$.}
\label{table:1}
\begin{center}
\begin{tabular}{ccc}
$l$ &$N_W(\widetilde{S})$ &$N_W(\widetilde{D})$\\
\hline
0& 0& 2 \\
2& 2& 0\\
$+1$& $+1$& $-1$\\
$-1$& $-1$& $+1$ \\
\end{tabular}
\end{center}
\end{table}
\end{minipage}
\begin{minipage}[t]{1cm}\hfill\end{minipage}
\begin{minipage}[t]{7.5cm}
of phase for the $d$-wave OP is $2\pi (l\pm 2)$.
Thus we obtain  winding numbers of $s$-wave and $d$-wave OPs in Table I. 
Thus we see that $s$-wave and $d$-wave have opposite winding numbers 
for the cases $l=+1$ and $l=-1$.
The opposite winding of $s$-wave and $d$-wave OPs has been pointed 
by  Volovik\cite{rf:Vol} in  an  isolated vortex, which corresponds 
to the case $l=-1$. 
Note that this relation holds also in the cases of vortex lattice for
$l=\pm 1$.
\end{minipage}
  
In conclusion, 
we obtained well-defined OPs at a site $(m,n)$: $\widetilde{S}(m,n)$, 
$\widetilde{D}(m,n)$, $\widetilde{P_x}(m,n)$ and $\widetilde{P_y}(m,n)$, 
which have a well-defined nature such that an  OP at the site obtained 
under translation  by a basis vector (of the vortex lattice) from a site 
$(m,n)$ is expressed by the corresponding OP at the site $(m,n)$ times 
a phase factor. Similar group theoretical analysis  also works 
for the case in which the center of the vortex is located in the middle 
of a plaquette of the  crystal lattice. \\

The authors would like to express their thanks to M. Ichioka
for  helpful discussions. 


\end{document}